\documentstyle[epsf,twoside,fleqn,espcrc2]{article}
\newcommand\eprint[1]{\publishedeprint{#1}{#1}}
\newcommand\publishedeprint[2]{\href{http://xxx.lanl.gov/abs/#2}{#1}}

\newcommand{\beq}{\begin{equation}}
\newcommand{\eeq}{\end{equation}}
\newcommand{\beqa}{\begin{eqnarray}}
\newcommand{\eeqa}{\end{eqnarray}}
\renewcommand{\ensuremath}[1]{\relax\ifmmode#1\else$#1$\fi}
\newcommand{\mnod}{\ensuremath{{\mQ^0}}}
\newcommand{\mpert}{\ensuremath{{\mQ^{pert}}}}
\newcommand{\mHH}{\ensuremath{{\mQ^{HH}}}}
\newcommand{\mQ}{\ensuremath{m_Q}}
\newcommand\0{\hphantom{0}}
\newcommand\ds{\displaystyle}
\newcommand\MeV{{\rm MeV}}
\newcommand\lsim
  {\mathrel{\lower5pt\hbox{$\buildrel{\textstyle<}\over{\textstyle\sim}$}}}
\newcommand\percent{%
\%
}

\title{$f_B$ and $f_{B_s}$ using NRQCD}

\author{T.~Bhattacharya\address{T-8 Group, MS B285, Los Alamos
  National Laboratory, Los Alamos, New Mexico 87545, U.S.A.}
  \thanks{In collaboration with A.~Ali~Khan (OSU), S.~Collins and
  C.T.H.~Davies, (Glasgow Univ., UKQCD Collab.), R.~Gupta (LANL),
  J.~Shigemitsu (OSU) and J.~Sloan (Univ. of Kentucky).}}

\begin{document}
\ifx\href\undefined\newcommand\href[2]{#2}\fi

\begin{abstract}
This talk summarizes the quenched calculation of $f_B$ and $f_{B_s}$ 
presented in~\cite{fBpaper}.  The heavy quark is simulated
using an $O(1/M^2)$ improved NRQCD action, and the tadpole improved
clover action is used for the light quarks.  The axial current
includes the $O(1/M)$ corrections in one-loop perturbation theory and
the $O(1/M^2)$ ones at tree-level.  We discuss the various systematic 
effects and consistency checks made in the calculation. 
%% find $f_B =152(11)({}^{+8}_{-12})(9)(6)$ MeV, $f_{B_s} =
%% 181(8)({}^{+7}_{-10})(11)(7)({}^{+7}_{-0})$ and $f_{B_s}/f_B =
%% 1.20(4)({}^{+4}_{-0})$, where, in each case, the first error is
%% statistical and the rest are various systematics.
\end{abstract}

\maketitle

\section{NRQCD ACTION AND OPERATOR}
The forward-backward symmetric evolution of the heavy quark propagator
$G_t$ using the NRQCD action~\cite{Lepage} is given by 
\begin{eqnarray}
G_{t+1} ={}&\ds \left(1-\frac{\delta^- H}{2}\right)
             \left(1-\frac{H_0}{2n}\right)^n \nonumber \\ 
           &\ds U_4^{\dagger}
             \left(1-\frac{H_0}{2n}\right)^n
             \left(1-\frac{\delta^+ H}{2}\right) G_t,
\end{eqnarray}
\iffalse
where $H_0$ and $\delta H$ are given by
\begin{eqnarray}
H_0 &=& -\frac{\Delta^{(2)}}{2\mnod},\\
\delta^{\pm} H &=& \ds -\frac{g\vec{\sigma}\cdot\vec{B}}{2\mnod}
                       +\frac{ig}{8\mnod^2}
                          \left(\vec\Delta\cdot\vec{E^{\pm}} -
                          \vec{E^{\pm}}\cdot\vec\Delta\right) \nonumber \\
               &  & \ds{} -\frac{g}{8\mnod^2}\vec{\sigma}\times
                          \left(\vec\Delta\times\vec{E^{\pm}} -
                          \vec{E^{\pm}}\times\vec\Delta\right) \nonumber \\
               &  & \ds{} -\frac{(\Delta^{(2)})^2}{8\mnod^3} + 
                        \frac{a^2\Delta^{(4)}}{24\mnod} -
                        \frac{a(\Delta^{(2)})^2}{16n(\mnod)^2}\,.
\label{eqn:ham}
\end{eqnarray}
\else 
where $H_0 = \Delta^{(2)}/{2\mnod}$ and $\delta^{\pm} H$ is defined in
\cite{fBpaper}.
\fi
The gauge links are tadpole improved by dividing them by $u_0 =
0.87779$.  The tree-level contribution $\exp(-\mnod t)$ of the bare
mass $\mnod$ to the heavy quark propagator is omitted from the
calculation.  The leading correction proportional to $\alpha_s$
redefines $\mnod$: we take this into account perturbatively when
matching the lattice theory to the continuum.

The current J, expanded in powers of $1/M$, is 
\begin{eqnarray}
  J &=& \eta_0 J^{(0)} + \frac{\eta_1}{2M} J^{(1)}  +
	    \frac{\eta_2}{2M} J^{(2)} \nonumber\\
    & &{} + \frac{\eta_3}{8M^2} J^{(3)} + \frac{\eta_4 g} {8M^2} J^{(4)} 
            + \frac{i\eta_5 g}{4M^2} J^{(5)}\,,
\label{eq:current}
\end{eqnarray}
where we keep all operators of dimension three and four, and only the 
three dimension five operators that appear at the tree level:
\begin{eqnarray}
J^{(0)} &=& \bar{q}\gamma_5\gamma_4 Q, \nonumber\\
J^{(1)} &=& \bar{q}\left(\vec{\gamma}\cdot\vec{\Delta}\right) 
			\gamma_5\gamma_4 Q,   \nonumber\\
J^{(2)} &=& \left(\vec{\Delta} \bar{q}\cdot\vec{\gamma}\right)
			\gamma_5\gamma_4 Q,   \nonumber\\
J^{(3)} &=& \bar{q} \Delta^2 \gamma_5\gamma_4 Q \nonumber\\
J^{(4)} &=& \bar{q} \vec{\Sigma}\cdot\vec{B} \gamma_5\gamma_4 Q \nonumber\\
J^{(5)} &=& \bar{q} \vec{\alpha}\cdot\vec{E} \gamma_5\gamma_4 Q\,.
\label{eq:currents}
\end{eqnarray}
$q$ is the four component light spinor and $Q$ is the nonrelativistic
heavy quark spinor whose upper two components are zero.

On transcribing $J$ on to the lattice, the mixing induced by the hard
cutoff has to be included. Dropping dimension five and higher
operators, the lattice currents $J_L$ can be written as
\begin{equation}
\newcommand\m[1]{\(\displaystyle#1\)}
\left(\vcenter{\hbox{{\m{J^{(0)}_L}}} \hbox{\m{J^{(1)}_L}} \hbox{\m{J^{(2)}_L}}
}\right)
=
\left(\vcenter{
\halign{&\hfil#\hfil\cr
\m{\zeta_{00}}       & \m{a\zeta_{01}} & \m{a\zeta_{02}} \cr
\m{a^{-1}\zeta_{10}} & \m{\zeta_{11}}  & \m{\zeta_{12}}  \cr
\m{a^{-1}\zeta_{20}} & \m{\zeta_{21}}  & \m{\zeta_{22}}  \cr
}}\right)
\left(\vcenter{\hbox{{\m{J^{(0)}}}} \hbox{\m{J^{(1)}}} \hbox{\m{J^{(2)}}}
}\right)\,.
\label{eq:latcurrent}
\end{equation}
In the final expression for $J$~\cite{perttheory}, every $a^{-1}$
appears multiplied by an $M^{-1}$; all infrared divergences cancel;
and the final logarithmic scale dependence is of the form
$\log{a\mnod}$.  Apart from these logarithms, all the coefficients are
finite for $1 \lsim a\mnod \le \infty$.  We shall highlight the contribution 
of the different currents $J^{(i)}$ to $f_B$. 

\section{LATTICE PARAMETERS}

The statistical sample consists of 102 $16^3\times 48$ quenched
lattices at $\beta = 6/g^2 = 6.0$.  Heavy quarks are simulated with
$a\mnod = 1.6$,$2.0$,$2.7$ with $n=2$; and $4.0$,$7.0$,$10.0$ with $n=1$.
Light quarks with $\kappa = $ 0.1369, 0.1375 and 0.13808 (bracketing
the strange quark) were simulated using the tadpole improved clover
action.  Details of the simulations 
%% (update of lattices, smearing
%% functions used as sources in the inversion of quark propagators, $etc.$) 
and fixing of the lattice parameters are given in \cite{fBpaper}.
Using the data for $m_\pi^2$ and $m_\rho$ versus $\kappa$ we get
$\kappa_l = 0.13917(9)$ and $a^{-1} = 1.92(7)$ GeV.  For the strange
quark mass we use $\kappa_s = 0.1376(1)$ from $m_K$ to quote the
central value, and $\kappa_s = 0.1372(3)$ from $m_{K^*}$ to estimate
the error.  Lastly, a reanalyses of the $B$ spectrum gives $aM_b^0 =
2.32(12)$~\cite{spectrumpaper}, whereas in \cite{fBpaper} we had used
$aM_b^0 = 2.22(11)$. This revised $aM_b^0$, however, leads to an
insignificant change in estimates of decay constants.  Estimates of
$f_{B_s}$, which only require an interpolation in $\kappa$ and
$\mnod$, are more reliable than those for $f_B$, which need a large
extrapolation in $\kappa$.

\section{METHOD OF ANALYSIS}

To estimate uncertainty in measuring the mass and amplitude of two point correlators, 
we extract the decay constants in three ways:
\begin{enumerate}
\item{} Individual local-smeared correlators $\langle J_i S\rangle$
        and the smeared-smeared correlators $\langle S S\rangle$ are
        fit using the form $A \exp{-Et}$.  From these we get the
        contribution of each bare operator to the decay constant: $
        f_i \sqrt{M} = {A_i}/{\sqrt{A_S}}$.  These $f_i \sqrt{M}$ are
        combined using Eq.~\ref{eq:current} to obtain the decay
        constant for each heavy-light combination.  Finally, we
        extrapolate/\discretionary{}{}{}interpolate in $\kappa$ and
        $a\mnod$.
\item{} We combine the $ J_i$  according to
	Eq.~\ref{eq:current} to obtain $\langle J S\rangle$.  We fit
	this and the $\langle S S\rangle$ to exponential forms $A
	\exp{-Et}$ and obtain the decay constant $f\sqrt{M}$ at each
	combination of heavy and light quarks.  These are then
	interpolated/extrapolated in $\kappa$ and $a\mnod$.
\item{} We make simultaneous fits to each $\langle J_i S\rangle$ and
        the $\langle S S\rangle$ over the same fit range to obtain the
        $f_i \sqrt{M}$.  We then extrapolate/interpolate these to the
        physical masses. Finally, we combine these using
        Eq.~\ref{eq:current} to obtain $f_B$ and $f_{B_s}$.
\end{enumerate}
We find that for $a\mnod = 1.6$, 2.0 and 2.7, all three methods
of analysis give the same result, whereas for $a\mnod = 4.0$ the
agreement is marginal.  Consequently, we use only the first three
values of $a\mnod$ to obtain our final results.

\begin{table}[t]
%\caption{$f_i \sqrt{M}$ from various operators versus $a\mnod$.}
\caption{Various contributions to \(f_Q\sqrt{M_Q}\) versus $a\mnod$.}
\label{tab:individual}
\begin{center}
\setlength\tabcolsep{0.7pt}
\begin{tabular}{|l|l|l|l|l|l|}
\hline
         \(a\mnod\)
         & \multicolumn{1}{c|}{\(\displaystyle{J^{(0)}}\)} 
	 & \multicolumn{1}{c|}{\(\displaystyle{J^{(1)}}\)}
	 & \multicolumn{1}{c|}{\(\displaystyle{J^{(3)}}\)}
	 & \multicolumn{1}{c|}{\(\displaystyle{J^{(4)}}\)} 
	 & \multicolumn{1}{c|}{\(\displaystyle{J^{(5)}}\)} \\
%% \(a\mnod\)&\multicolumn{1}{c|}{(0)}
%%      &\multicolumn{1}{c|}{(1)}
%%        &\multicolumn{1}{c|}{(3)}
%%        &\multicolumn{1}{c|}{(4)}
%%        &\multicolumn{1}{c|}{(5)}\\
\hline
1.6 & .164(7)  & -.025(1) & -.0053(4) & .0036(2) & -.0050(3) \\
2.0 & .167(9)  & -.021(1) & -.0036(2) & .0022(1) & -.0033(2) \\
2.7 & .170(11) & -.016(1) & -.0022(1) & .0011(1) & -.0018(1) \\
4.0 & .174(11) & -.011(1) & -.0011(1) & .0005(0) & -.0008(1) \\
\hline
\end{tabular}
\vskip-2\baselineskip
\end{center}
\end{table}

Estimates of $f_i\sqrt{M}$, multiplied by their tree-level
coefficients and extrapolated to $\kappa_l$, are presented in
Table~\ref{tab:individual}.  We expect that the bare matrix element of
$J^{(1)}/2\mnod$ should be smaller than that of $J^{(0)}$ by a factor
of $O(\alpha/a\mnod) \sim O(\Lambda_{QCD}/\mnod)$, and qualitatively
this is borne out by the data.  Similarly, the bare matrix elements of
$J^{(3)}/8(\mnod)^2$, $J^{(4)}/8(\mnod)^2$ and $J^{(5)}/4(\mnod)^2$
are smaller than that of $J^{(1)}$ by a factor of $O(1/a\mnod)$.

\section{RESULTS}
\begin{table}[t]
\caption{$f_Q\sqrt{M_Q}$ extrapolated to $\kappa_l$ and $\kappa_s$. Results 
are given for $aq^\ast=1$ and $\pi$.}
\label{tab:vsheavy}
\setlength\tabcolsep{3pt}
\begin{center}
\vspace{\baselineskip}
\begin{tabular}{|l|l|l|l|l|l|}
\hline
         & \multicolumn{2}{c|}{\(\kappa_l\)}
         & \multicolumn{2}{c|}{\(\kappa_s\)}\\
\cline{2-3} \cline{4-5}
\(a\mnod\) & \multicolumn{1}{c|}{\(1/a\)}        
         & \multicolumn{1}{c|}{\(\pi/a\)}
         & \multicolumn{1}{c|}{\(1/a\)}        
         & \multicolumn{1}{c|}{\(\pi/a\)} \\
\hline
%%\(a\mnod\) & {\(\kappa_l,1/a\)}      & {\(\kappa_l,\pi/a\)}
%%         & {\(\kappa_s,1/a\)}  & {\(\kappa_s,\pi/a\)} \\
1.6 &     {0.114(6)} &     {0.121(6)}
    & {0.136(4)} & {0.144(4)} \\
2.0 &     {0.121(7)} &     {0.128(7)}
    & {0.144(5)} & {0.153(5)} \\
2.7 &     {0.127(9)} &     {0.135(9)}
    & {0.153(6)} & {0.164(6)} \\
4.0 &     {0.135(9)} &     {0.144(10)}
    & {0.164(6)} & {0.176(6)} \\
\hline
\end{tabular}
\vskip-2\baselineskip
\end{center}
\end{table}

Table~\ref{tab:vsheavy} presents $f_Q\sqrt{M_Q}$ extrapolated to
$\kappa_l$ and $\kappa_s$ for two choices of the lattice scale we
believe covers the range relevant to this calculation, $aq^\ast = 1,\
\pi$. (The lattice coupling at these scales are $\alpha_V(\pi/a) =
0.1557$ and $\alpha_V(1/a) = 0.2453$.)  For our final values we
average the two estimates and interpolate to $aM_b^0$:
\begin{eqnarray}
f_B     &=& 147(11)({}^{+8}_{-12})(09)(6) \MeV\nonumber\\
f_{B_s} &=& 175(08)({}^{+7}_{-10})(11)(7)({}^{+7}_{-0}) \MeV\nonumber\\
\frac{f_{B_s}}{f_B} &=& 1.20(4)({}^{+4}_{-0})\,.
\end{eqnarray}

The error estimates are obtained as follows. To determine the
uncertainty due to setting the lattice scale, we repeat the
calculation with $a^{-1} = 1.8$ and $2$ GeV~\cite{fBpaper}.  This
gives a variation $({}^{+8}_{-12})$ MeV in $f_B$ and $({}^{+7}_{-10})$
in $f_{B_s}$.

To estimate the perturbative errors we consider three issues. (i) The
results at $q^\ast = 1$ and $\pi$ differ by about $5\percent$.  (ii)
The dimension five operators have been included with tree-level
coefficients and their mixing with lower dimensional operators
neglected.  This, we estimate, introduces an error of
$O(\alpha_s/(Ma)^2) \sim 4\percent$.  (iii) The neglected higher
dimension operators are expected to be $O(\Lambda_{QCD}/M)^3 <
1\percent$.  Altogether we assign a $\sim 7\percent$ error coming from
these sources.

The remaining discretization errors are $O(\Lambda_{QCD}a)^2$,
which we estimate to be $4\percent$. 
The last error in $f_{B_s}$ and $f_{B_s}/f_B$ is due to the uncertainty in 
$\kappa_s$. We use the difference in $\kappa_s$ obtained from $K$ and $K^*$ 
to estimate this. 

An alternate approach is to neglect the contribution of the dimension
five operators $J^{(3)}$, $J^{(4)}$ and $J^{(5)}$ in the final result, and 
use them only to estimate errors.  The reason is that the neglected 
mixing with $J^{(0)}$ could introduce $O(\alpha_s/(Ma)^2)$ errors, which 
can be larger than their $O(\Lambda_{QCD}/M)^2$ contribution.  Neglecting 
the dimension five operators (in which case the calculation is consistent 
to $O(1/M)$ and $O(\alpha/M)$) would give 
\begin{eqnarray}
f_B     &=& 152(11)({}^{+8}_{-12})(10)(6) \MeV\nonumber\\
f_{B_s} &=& 181(08)({}^{+7}_{-10})(12)(7)({}^{+7}_{-0}) \MeV \,.
\end{eqnarray}

Finally, we give a breakup of the various contributions.  The
tree-level estimate is $f_B = 169$ MeV. Of this $195$ MeV comes from
$J^{(0)}$, $-21$ MeV from $J^{(1)}$, and $-5$ MeV from the $1/M^2$
corrections.  The one-loop corrections reduce the estimates by
$\sim13\percent$.  On the other hand we find that the ratio
$f_{B_s}/f_B$ is insensitive to the various corrections.

\begin{table}[t]
\caption{Comparison with other lattice determinations. The three
groups of results use NRQCD, Fermilab and Wilson or Clover formalisms
respectively.  The errors are statistical and systematic (excluding quenching) 
combined in quadrature.}
\label{tab:final}
\setlength\tabcolsep{3pt}
\begin{center}
\begin{tabular}{|l|l|l|l|}
\hline
 &\multicolumn{1}{c|}{\(f_B\)}
 &\multicolumn{1}{c|}{\(f_{B_s}\)}
 &\multicolumn{1}{c|}{\(f_{B_s}/f_B\)}\\
\hline
\cite{fBpaper}
          &\( 147(11)(^{+13}_{-16}) \)
          &\( 175(8)(16)          \)
	  &\( 1.20(4)(^{+4}_{-0}) \)\\
\cite{JLQCDNRQCD}
          &\( 162(7)(^{+34}_{-16}) \)
          &\( 190(5)(^{+42}_{-18}) \)
	  &\( 1.18(3)(5) \)\\
%% UKQCD  &\(  160(6)(^{+53}_{-19})  \)
%%        &\( 194(^{+6}_{-5})(^{+62}_{-9} \))
%% 	  &\( 1.22(^{+4}_{-3}) \)\textBlack\\ 
\hline
\cite{Fermilab}
          &\(  164(^{+14}_{-11})(8)  \)
          &\( 185(^{+13}_{-8})(9) \)
	  &\( 1.13(^{+5}_{-4}) \)\\
\cite{JLQCDFermilab}
          &\(  173(4)(12)             \)
          &\( 199(3)(14)           \)
	  &\( \)\\
\hline
\cite{APE}
          &\(  180(32)               \)
          &\( 205(15)(31)         \)
          &\(  1.14(8)               \)\\
\cite{MILC}
          &\(  159(11)(^{+22}_{-9}) \)
          &\( 175(10)(^{+28}_{-10}) \)
	  &\( 1.11(2)(^{+4}_{-3}) \)\\
\hline
\end{tabular}
\vskip-2\baselineskip
\end{center}
\end{table}

\section{DISCUSSIONS}

In Table~\ref{tab:final} we compare our results with other
calculations.  The formalisms used for incorporating heavy quarks are
NRQCD~\cite{JLQCDNRQCD}, Fermilab~\cite{Fermilab,JLQCDFermilab},
Wilson/clover~\cite{APE}, and Wilson~\cite{MILC}. Within errors the
different approaches give consistent results for the quenched theory.
Further improvement of the NRQCD estimates needs the full operator
mixing matrix at $O(1/M^2)$, which in turn requires improving the
light quark action to $O(a^2)$.  A necessary check of our results is
to show that they are independent of $a$.  We are addressing these
issues.

A surprising feature of these data is that the ratio $f_{B_s}/f_B$,
which is insensitive to most of the systematic errors, shows a
difference between the NRQCD and the other formalisms. We do not 
understand this. Possible issues to investigate are (i) the 
chiral extrapolation and (ii) the dependence on the heavy 
quark mass as this is handled differently in the different approaches.

\subsection*{Acknowledgements}
We acknowledge the support of the ACL at LANL and NCSA at Urbana Champagne.

\end{document}